\documentclass[amsmath, amssymb, aps, pra,twocolumn,floatfix,nofootinbib]{revtex4}
\usepackage[utf8]{inputenc}
\usepackage{titlesec}
\usepackage[colorlinks]{hyperref}
\usepackage{amsmath,amssymb}
\usepackage{bbold}
\usepackage{float}
\usepackage{dcolumn}
\usepackage{natbib}
\usepackage{bm}
\usepackage{dsfont}
\usepackage{url}
\usepackage{mathtools}
\usepackage{amsthm}
\usepackage{amsfonts}
\usepackage{braket}
\usepackage{cancel}
\usepackage{bbold}
\usepackage{makecell}
\usepackage[normalem]{ulem}
\usepackage[dvipsnames]{xcolor}
\usepackage{comment}

\usepackage{afterpage}
\usepackage{multirow}
\usepackage{array}

\usepackage[toc,page]{appendix}

\begin{document}
\title{Quantum subsets discrimination and key sifting with success probability one}
\author{D.A.\,Kronberg}

\affiliation{Steklov Mathematical Institute, Russian Academy of Sciences, 8 Gubkina Street, Moscow 119991, Russia
\\ dmitry.kronberg@gmail.com
}

\begin{abstract}
Non-distinguishability of non-orthogonal quantum states is the essential feature of quantum mechanics. In particular, one cannot deterministically orthogonalize the set of quantum states, while probabilistic orthogonalization is possible, including unambiguous state discrimination and discrimination between the subsets. Here, we propose a method for error-free subsets discrimination with significantly increased success probability, which may reach one even for non-orthogonal and linearly dependent input states. This method may be applied in quantum cryptography: An alternative basis reconciliation procedure is described, which may provide unit sifting probability even for a large number of bases and for non-orthogonal states within each basis.
\end{abstract}

\maketitle

Some restrictions of quantum mechanics do not work, if postselection is allowed, i.e., when there is a failure outcome, and corresponding attempts may be discarded. Probabilistic cloning \cite{probabilistic_cloning} or unambiguous state discrimination (USD) \cite{Ivanovic_USD, Dieks_USD, Peres_USD} are well-known examples of postselective operations, which in case of success bypass the corresponding restrictions.

A well-studied problem is finding the maximal success probability for some postselective operations, including USD \cite{Bergou_optimal_USD} and filtering \cite{bergou_filtering}, or, more generally, quantum subsets discrimination \cite{subsets_discrimination}. For non-orthogonal quantum states, this probability is always below one, which displays uncertainty inherent in quantum mechanics.

However, as we demonstrate below, this uncertainty may sometimes not interfere with overall success probability of the operation. Specifically, we show that if the type of success is unpredictable, the success probability may become high, and can even reach one. Hence, we bypass the restriction by accepting unpredictability, which follows the spirit of quantum mechanics.

A key observation about quantum operations on pure states is that they can be associated with convex combinations of the Gram matrices (i.e., matrices of pairwise scalar products):
\begin{equation}
\label{eq:decomposition_corollary}
G_{\rm in} = \sum_k p_k G_k.
\end{equation}
For any such combination, the set of pure states with the Gram matrix $G_{\rm in}$ can be mapped onto several possible sets of pure states, with each set having Gram matrix $G_k$, and with mapping probability $p_k$. See Appendix A for more detailed description and for the construction of the corresponding Kraus map.

Now, as a quick example of the main idea, consider the Gram matrix
\begin{equation}
G_{\rm in} = \begin{pmatrix}
1 & \frac 13 & \frac 13 \\
\frac 13 & 1 & \frac 13 \\
\frac 13 & \frac 13 & 1
\end{pmatrix}
\end{equation}
of three non-orthogonal equidistant states $\{|\varphi_i\rangle\}_{i=1}^3$, with $\langle\varphi_i|\varphi_j\rangle = 1/3$ for $i\neq j$, and the
following decomposition:
\begin{eqnarray}
G_{\rm in} = p_{(1:23)}G_{(1:23)} + p_{(2:13)} G_{(2:13)} + p_{(12:3)} G_{(12:3)} \nonumber\\
= \frac 13 \begin{pmatrix}
1 & 0 & 0 \\
0 & 1 & 1 \\
0 & 1 & 1
\end{pmatrix} + \frac 13 \begin{pmatrix}
1 & 0 & 1 \\
0 & 1 & 0 \\
1 & 0 & 1
\end{pmatrix} + \frac 13 \begin{pmatrix}
1 & 1 & 0 \\
1 & 1 & 0 \\
0 & 0 & 1
\end{pmatrix}.
\label{eq:example_decomposition}
\end{eqnarray}
Here, there are three possible output options, which correspond to three subsets divisions. For the first division, $(1:23)$, the operation is the map of $\{|\varphi_1\rangle, |\varphi_2\rangle, |\varphi_3\rangle\}$ on three states $\{|e_0\rangle, |e_1\rangle, |e_1\rangle\}$, where two last states do coincide, and $\langle e_0|e_1\rangle = 0$. Hence, this operation discriminates between the state $|\varphi_1\rangle$, and the subset $\{|\varphi_2\rangle, |\varphi_3\rangle\}$ without error, because two output states are orthogonal, and measurement in $\{|e_0\rangle, |e_1\rangle\}$ basis identifies the subset. Such error-free discrimination of one state from the others is also referred as quantum state filtering \cite{bergou_filtering}. Two other divisions, $(2:13)$ and $(12:3)$, corresponds to filtering of two other states, $|\varphi_2\rangle$ and $|\varphi_3\rangle$, or, equivalently, discrimination between two other subsets.

The remarkable thing is that \emph{there is no failure for the corresponding operation}. Any option corresponds to successful filtering, thus success probability is one, but quantum uncertainty is that one cannot predict which subsets will be discriminated.

Previous studies were focused on maximizing success probability for one particular division, and this probability can be done higher than $1/3$, but still does not reach one. The situation changes dramatically when several divisions are taken into account. Similarly, the proposed operation can also be viewed as a variant for non-optimal unambiguous state discrimination (USD), because each state is discriminated from the others with success probability $1/3$, which is below optimal USD success probability $2/3$ for this configuration. From this point of view, in our construction, we sacrifice success probability for a given partition, so that the set of states in case of failure allows for additional filtering attempt, now for other partitions. This approach resembles quantum information recycling idea \cite{recycling}. Thus, quantum information recycling for pure states is simply a further decomposition of the failure Gram matrix, and the observation is that non-optimized postselective transformations or measurements may lead to effective operation including recycling.

Therefore, the key idea is, that if we accept unpredictability of the division, success probability for subsets discrimination may be increased, and for some cases, including the above example, it may reach one. Mathematically, it is the decomposition of the input Gram matrix $G_{\rm in}$ to the Gram matrices $G_{(L:R)} \geq 0$ with $g_{ii} = 1$ and with zeros on the divisions intersections:
\begin{equation}
\label{eq:zero_elements}
    g_{ij} = 0,\ {\rm if}\ (i\in L, j\in R)\ {\rm or}\ (i\in R, j\in L).
\end{equation}

We take into account non-equal probabilities of a particular $(L:R)$ division for input states by using diagonal matrices $P_{(L:R)} = {\rm diag}(\{\sqrt{p((L:R)\,|\,i)}\}_i)$ (see Appendix A). There may also be a residual matrix $G_f$ which corresponds to overall failure, with similar conditional probabilities matrix $P_f$, and the general procedure is decomposition of $G_{\rm in}$ into a convex combination
\begin{equation}
\label{eq:decomposition_main}
G_{\rm in} = \sum_L P_{(L:R)}G_{(L:R)}P_{(L:R)} + P_f G_f P_f,
\end{equation}
which can be further generalized to include higher number of subsets for non-binary discrimination. E.g., the division $(1:2:3:...:N)$ corresponds to the USD.

Average success probability for \textit{a priori} state probabilities $\{p_i\}_{i\in X}$ and for the set $Y_C$ of conclusive results reads
\begin{equation}
    p_{\rm success} = \sum_{i\in X, k \in Y_C}p_i p(k\,|\,i),
\end{equation}
where summation is over all conclusive outcomes. For subsets discrimination, these outcomes are labeled as $(L:R)$, and for uniform probability distribution success probability reads
\begin{equation}
p_{\rm success} = \sum_{i, (L:R)}\frac{p((L:R)\,|\,i)}{N} = \frac 1N{\rm Tr}\left(\sum_LP_{(L:R)}^2\right).
\end{equation}
Thus, the proposed randomized measurement allows making this success probability high, if the unpredictability of the division is satisfactory.

Note that decomposition of the form (\ref{eq:decomposition_main}) may be not unique, and different $P_{(L:R)}$ matrices may lead to the same sum, as well as one may chose different non-diagonal elements for each $G_{(L:R)}$, when these elements are non-zero. It allows to adjust this procedure, for example, for maximization of the probability for the desired set of divisions.

For a given Gram matrix, it may be difficult to find the decomposition of form (\ref{eq:decomposition_main}) with minimal or zero failure probability, partly due to the non-uniqueness of possible partitions and probabilities. Nevertheless, for particular cases, this decomposition is relatively easy, and this approach allows for the composition of the right-hand side matrices into the final result.

As an example, consider even $N$ and balanced divisions $(L:R)$ with equal number of elements, $|L| = |R|$. There are $C_{N-1}^{N / 2 - 1}$ of such divisions, where $C_N^K = N!/K!(N-K)!$ is binomial coefficient. For all divisions, let us consider Gram matrices with $g_{ij} = 1$ if $i, j \in L$ or $i, j \in R$. The sum of corresponding Gram matrices will have $C_{N-1}^{N / 2 - 1}$ on the diagonal, and $C_{N-2}^{N / 2 - 2}$ outside the diagonal. Let us set the same probability $p_k = 1/C_{N-1}^{N / 2 - 1}$ for each division, and the sum (\ref{eq:decomposition_corollary}) will result to the Gram matrix $G_{\rm in}$ of equidistant states $\{|\varphi_i\rangle\}_{i=1}^N$ with 
\begin{equation}
\label{eq:equidistant_product}
\langle\varphi_i|\varphi_j\rangle = \frac{C_{N-2}^{N / 2 - 2}}{C_{N-1}^{N / 2 - 1}} = \frac{N - 2}{2N - 2},\ \forall i\neq j.
\end{equation}

Hence, this method allows to perform conclusive subsets discrimination for equidistant states, if $\langle\varphi_i|\varphi_j\rangle = c = (N - 2) / (2N - 2)$. For larger inner product values $\langle\varphi_i|\varphi_j\rangle = \zeta$, residual all-ones matrix adds inconclusive result probability $p_f = (\zeta - c) / (1 - c)$.

As a possible optical implementation for $N$ equidistant states, let us consider $N$-mode coherent states of the form $\{|\alpha\rangle|0\rangle...|0\rangle, |0\rangle|\alpha|0\rangle ... |0\rangle, |0\rangle|0\rangle|\alpha\rangle...|0\rangle, |0\rangle|0\rangle|0\rangle...|\alpha\rangle\}$ with real $\alpha$. Inner product between different states equals $\zeta = e^{-\mu}$, where $\mu = \alpha^2$ is the intensity. USD probability for this set of states equals $1 - e^{-\mu}$, which is probability of finding at least one photon, and its mode determines the state. But error-free subsets discrimination probability is larger by a factor of $(1 - p_f)/(1 - \zeta) = 1/(1 - c) = 2(N - 1)/N$, this factor approaches $2$ for large number of states $N$. If $e^{-\mu} = (N-2)/(2N-2)$, subsets discrimination is possible with probability one, while non-vacuum probability is lower. E.g., for $N=4$, conclusive subsets discrimination is possible for $\mu\approx 1.1$ photons per pulse, and non-vacuum probability is $2/3$. For $N=32$, $\mu\approx0.73$ allows for conclusive subsets discrimination, and non-vacuum probability is $\approx 0.516$. Hence, remarkably, error-free information can be obtained with considerably larger probability than probability of finding a photon.

The task of subsets discrimination can be viewed as a particular case of quantum state exclusion (QSE) task \cite{conclusive_exclusion, exclusion_group_generated}. In QSE, each outcome corresponds to eliminating some of states, and for subsets discrimination, the states of the remaining subset are eliminated. USD is also a particular case of QSE, where all states except for one are excluded. The most studied case is one-state exclusion, including mixed states at input. If one-state exclusion is possible for every state without inconclusive outcome (conclusive QSE), this set is called antidistinguishable \cite{heinosaari2018antidistinguishability, antidistinguishability2025}, and this property plays an essential role in ontological properties of quantum states \cite{pusey2012reality}.

The proposed method of Gram matrix decomposition (\ref{eq:decomposition_lemma}) is suitable for QSE, while the output matrices have other form. Single state exclusion imposes the requirement $p(k\,|\,k) = 0$, thus $k$-th row and $k$-th column of $P_kG_kP_k$ in (\ref{eq:decomposition_lemma}) are all-zeros. Hence, antidistinguishability is formalized in terms of Gram matrix decomposition, which duplicates Theorem 3.2 in Ref. \cite{antidistinguishability2025} without using semi-definite programming (SDP) technique, purely relying on the properties of quantum states transformations. Similarly, decomposition (\ref{eq:decomposition_main}) describes particular case of $m$-ary QSE without relying on SDP, unlike its conventional study \cite{conclusive_exclusion}. Thus, this method significantly simplifies the study for some sets of pure states, while SDP is suitable for general case and for mixed states.

Quantum states discrimination is a core task for quantum information, hence there are a lot of potential applications of this result, despite its probabilistic nature. Let us describe a quantum cryptography application, where randomness is aligned with basis choice.

Conventionally in QKD, legitimate users use random independent basis choice, thus for large number of bases their matching probability becomes low, which restricts from using large number of bases. Let us propose an alternative basis reconciliation procedure, where the proposed method increases sifting probability. Remarkably, sifting probability can become one even for large number of bases with non-orthogonal states.

Consider a set of quantum states, for which conclusive discrimination of balanced subsets is possible, say, the equidistant states configuration with inner products (\ref{eq:equidistant_product}). Let Alice send a random state $|\varphi_j\rangle$ from this set for every message, there is no `basis choice' for Alice any more. Now, Bob performs the proposed discrimination, and obtains one of the divisions, $(L:R)$, and the particular set, say, $R$. Thus, Bob knows that $j \in R$. Now let Bob pick two random indices, $m\in L$, and  $n \in R$. Then, Bob announces whether $\{0:n,\ 1:m\}$ or $\{0: m,\ 1:n\}$, chosen at random. If $j \in \{m, n\}$, Alice agrees, and they use this announcement as a basis. Otherwise Bob selects two other indices, and he makes new announcements, until Alice agrees. As a result, they have a basis choice, and one of the states in this basis was sent by Alice. With this procedure, they will always agree on the basis, hence there is no more basis mismatch problem for them, despite a large number of possible bases, namely, $N(N-1)/2$. And, remarkably, conclusive probability for each basis is also one with this method, despite the non-orthogonality of the states.

Additionally, equidistant states configuration can be improved by making the states linearly dependent, to resist USD attack. For that, Alice and Bob can use Gram matrices with $g_{jk} = e^{i\theta}$ for $j>k$ and $g_{jk} = e^{-i\theta}$ for $j<k$, where $\theta$ satisfies
\begin{equation}
\label{eq:LD_criteria}
\frac{\sin\frac{\pi - \theta}{N}}{\sin(\theta + \frac{\pi - \theta}{N})} = \frac{N - 2}{2N - 2}.
\end{equation}
In this case, as shown in Ref.\cite{equidistant}, the resulting matrix $G_{\rm in}$ becomes degenerate, which implies linear dependence for the corresponding vectors.

We therefore note that linear independence is not a necessary condition for the proposed construction, and linearly dependent set of equidistant states can also provide success probability one for subsets discrimination.

For large number of states and for non-binary divisions, say, $(M_1:M_2:...:M_{2^l})$, the situation is remarkable. With the proposed communication, Alice and Bob may extract $l$ bits of key from their set of states, with limited eavesdropper's information due to non-orthogonality, and because there is no basis at the moment of attack.

The described basis reconciliation procedure is given for a model QKD protocol, described by states configuration and POVM. Here we do not propose its cost-effective implementation, based, for example, on coherent states and avalanche detectors. We also do not provide a complete security proof with estimating eavesdropper's information for arbitrary error. But let us show that this is a valid QKD protocol, where any eavesdropping attempt causes error and therefore can be detected. Two statements justify this claim:
\begin{enumerate}
    \item \textit{Any non-trivial measurement for non-orthogonal equidistant states introduces disturbance.} It follows from non-orthogonality. Hence, any eavesdropping attempt would damage the states.
    \item \textit{For any damaged quantum state, the proposed randomized measurement causes error.} Without loss of generality, consider original state $|\varphi_0\rangle$, and its damaged version, $\sum_ic_i|\varphi_i\rangle$, with $|c_0| < 1$. There exists $k\neq 0: |c_k| > 0$, hence there is at least one measurement operator $M_{(L:R)}$, for which $0 \in L, k \in R$, or vice versa. This operator would yield nonzero error.
\end{enumerate}

Hence, for zero error, one might be confident, that there is no information leakage to Eve. Therefore, the proposed states configuration and measurement allows for secret key generation with sifting probability one.

Previously it was widely believed that QKD should use whether random basis choice (with nonunit mismatch probability), or nonzero inconclusive result probability, if it uses non-orthogonal quantum states. With the proposed sifting procedure, large number of bases with non-orthogonal states still allow for sifting probability one.

The proposed method for subsets discrimination involves pure states, because their configuration allows for Gram matrix description, which is not the case for mixed states. Hence, application of this method for mixed states requires additional in-depth study. But still, the idea that several outcomes of the Kraus map may be useful, is also relevant for mixed states, while finding the Kraus operators for the desired transformation may be more difficult.

To conclude, it was shown that error-free subsets discrimination for the set of non-orthogonal quantum states may have success probability one, if the subsets division is unpredictable. Thus, the long-standing prohibition of error-free discrimination for non-orthogonal quantum states is not absolute. By relaxing the focus on one particular subsets division, we propose a method of randomized quantum state discrimination which can provide unit sifting probability in quantum cryptography.

This work was performed at the Steklov International Mathematical Center and supported by the Ministry of Science and Higher Education of the Russian Federation (agreement no. 075-15-2025-303).

\appendix

\section*{Appendix A: Gram decomposition}

Let us provide a mathematical tool which simplifies working with postselective transformations of pure quantum states. 
Quantum transformations (channels) may be described with Kraus operators $\{\hat M_k\}_{k=1}^M$, where $\sum_k \hat M_k^\dagger \hat M_k =I$. When the transformation is applied to a pure state $|\varphi\rangle$,  probability of the outcome $k$ is $p_k = {\rm Tr}\hat M_k|\varphi\rangle\langle\varphi|\hat M_k^\dagger$), and output state after $k$ reads
\begin{equation}
|\psi_k\rangle = \frac{1}{\sqrt{p_k}} \hat M_k|\varphi\rangle.
\end{equation}

For the set $\{|\varphi_i\rangle\}_{i=1}^N$ of pure states, their configuration is characterized by the Gram matrix
\begin{equation}
G = \{\langle \varphi_i|\varphi_j\rangle\}_{i, j = 1}^N.
\end{equation}

Let us now show that transformation of pure quantum states can be described as convex combination of the output Gram matrices.

\textbf{Lemma.} For every decomposition
\begin{equation}
\label{eq:decomposition_lemma}
G_{\rm in} = \sum_k P_k G_kP_k,
\end{equation}
there is a quantum channel which maps the set $\{|\varphi_i\rangle\}_{i=1}^N$ of input states to each set $\{|\psi_i^{(k)}\rangle\}_{i=1}^N$ of output states with the Gram matrix $G_k$, and the set of conditional probabilities is $\{p(k\,|\,i)\}_{ik}$, where the matrix $P_k = {\rm diag}(\{\sqrt{{p(k\,|\,i)}}\}_{i=1}^N)$.

\begin{proof} It is a simple consequence of Theorem 1 in Ref. \cite{kronberg2022success}, see also Theorem 2 in Ref. \cite{Chefles2004_existence}. For each outcome $k$, let us define operators
\begin{equation*}
\hat A = \sum_{i = 1}^R|\varphi_i\rangle\langle e_i|, \quad 
\hat B_k = \sum_{i = 1}^R|\psi_i^{(k)}\rangle\langle e_i|,
\end{equation*}
where $R$ is rank of $G_{\rm in}$, and $\{|e_i\rangle\}_{i=1}^R$ is orthonormal basis. Without loss of generality, let $\{|\varphi_i\rangle\}_{i=1}^R$ be linearly independent, hence there exists $\hat A^{-1}: \hat A^{-1}|\varphi_i\rangle = |e_i\rangle, \forall i \leq R$. Let us also introduce operators $\hat P_k = \sum_{i=1}^R\sqrt{{p(k\,|\,i)}}|e_i\rangle\langle e_i|$, $\hat G_{\rm in} = \sum_{i, j = 1}^R|e_i\rangle\langle \varphi_i|\varphi_j\rangle\langle e_j|$, $\hat G_k =\hat B_k^\dagger\hat B_k$. Now observe that
$$
\hat B_k \hat P_k \hat A^{-1}|\varphi_i\rangle = \sqrt{p(k\,|\,i)}|\psi_i^{(k)}\rangle\quad \forall i, k,
$$
thus, $\hat M_k =\hat B_k \hat P_k \hat A^{-1}$ performs the required map, and
\begin{eqnarray*}
\sum_k\hat M_k^\dagger\hat M_k = \sum_k (\hat A^{-1})^\dagger\hat P_k\hat B_k^\dagger \hat B_k \hat P_k \hat A^{-1} \\= \sum_k (\hat A^{-1})^\dagger\hat P_k \hat G_k \hat P_k\hat A^{-1} = (\hat A^{-1})^\dagger \hat G_{\rm in} \hat A^{-1}\\
= (\hat A^{-1})^\dagger \hat A^\dagger\hat A \hat A^{-1} = \hat I.
\end{eqnarray*}
Hence, the set $\{\hat M_k\}_k$ of Kraus operators specifies the corresponding channel.
\end{proof}

\textbf{Corollary.} If conditional probabilities $\{p(k\,|\,i)\}_{k}$ are all equal to each other for any $i$, matrices $P_k$ may be replaced by a scalars $\sqrt{p_k}$, and the decomposition (\ref{eq:decomposition_lemma}) becomes (\ref{eq:decomposition_corollary}).

\bibliography{references}

\end{document}